\documentclass{PoS}
\usepackage{graphicx, amssymb, amsfonts, amsmath, subfigure}

\title{The strong coupling regime of twelve flavors QCD}

\ShortTitle{The strong coupling regime of twelve flavors QCD}

\author{\speaker{Tiago Nunes da Silva}\\
        University of Groningen\\
        E-mail: \email{t.j.nunes@rug.nl}}
\author{Elisabetta Pallante\\
        University of Groningen\\
        E-mail: \email{e.pallante@rug.nl}}
\abstract{
We summarize the results recently reported in Ref.~\cite{Deuzeman:2012ee} [A.~Deuzeman, M.~P.~Lombardo, T.~Nunes da Silva and E.~Pallante,``The bulk transition of QCD with twelve flavors and the role of improvement''] for the $SU(3)$ gauge theory with $N_f=12$ fundamental flavors, and we add some numerical evidence and theoretical discussion. In particular, we study the nature of the bulk transition that separates a chirally broken phase at strong coupling from a chirally restored phase at weak coupling. When a non-improved action is used, a rapid crossover is observed at small bare quark masses. Our results  confirm a first order nature for this transition, in agreement with previous results we obtained using an improved action. As shown in Ref.~\cite{Deuzeman:2012ee}, when improvement of the action is used, the transition is preceded by a second rapid crossover at weaker coupling and an exotic phase emerges, where chiral symmetry is not yet broken. This can be explained \cite{Deuzeman:2012ee} by the non 
hermiticity of the improved lattice Transfer matrix, arising from the competition of nearest-neighbor and non-nearest neighbor interactions, the latter introduced by improvement and becoming increasingly relevant at strong coupling and coarse lattices. We further comment on how improvement may generally affect any lattice system at strong coupling, be it graphene or non abelian gauge theories inside or slightly below the conformal window.
}
\FullConference{The 30th International Symposium on Lattice Field Theory\\
                 June 24 - 29,  2012\\
                 Cairns, Australia}

\begin{document}

\section{Introduction}
The last years have seen a considerable effort being put into studies of (quasi) conformality on the lattice. These have been motivated not only by the theoretical interest in unravelling the phase diagram of gauge theories, but also by the possibility that (quasi) conformal strongly coupled dynamics plays a role in the description of physics beyond the Standard Model. 

Several recent lattice studies have focused on theories with many flavors of fermions in the fundamental representation, not only looking for the emergence of (quasi)conformality and its consequences in the continuum limit, but also investigating whether the conformal window scenario \cite{Miransky:1996pd} is realized. In this scenario, theories living in a range of number of flavors before the loss of asymptotic freedom are chirally restored and deconfined even at zero temperature. Each of these theories has an infrared fixed point (IRFP) where the theory is conformal. 

A key ingredient to identify theories below the conformal window is the occurrence of a finite temperature (thermal) transition from a hadronic (confined and chirally broken) to a Quark-Gluon Plasma (deconfined and chirally restored) phase. Inside the conformal window, a chiral symmetry breaking transition at zero temperature can still occur on the lattice. This transition, so called {\it bulk}, has no thermal behavior. It can be second order and signal the appearance of a UVFP (thus a new theory in the continuum limit) or it can be first order and have no continuum limit. A bulk transition was first observed in \cite{Damgaard:1997ut} for $SU(3)$ with $N_f=16$ fundamental flavors. It was later observed \cite{Deuzeman:2009mh,Deuzeman:2011pa} for the $N_f=12$ theory. Recently, the authors of \cite{deForcrand:2012vh} observe the line of bulk transitions to end for $N_f \simeq 51$, contrary to some expectations. 

Interestingly, we have also observed \cite{Deuzeman:2011pa} that, at sufficiently small bare quark masses and with an improved lattice action, the chiral symmetry breaking transition is preceded by a new phase of the system, signalled by a change of slope in the mass dependence of the chiral condensate \cite{Deuzeman:2012ee}. 
This phase was then also observed by \cite{Cheng:2011ic,Schaich:2012fr}, where it is also noted that the shift symmetry $S_4$ of staggered fermions is broken in this phase. It was recently shown in \cite{Deuzeman:2012ee} that the emergence of this phase is actually due to the improvement of the lattice action, and an explanation of its causes and effects has been attempted. 

In this proceedings we report on the nature of the chiral symmetry breaking bulk transition of $N_f=12$ with the non-improved lattice action, and further comment on the interplay of improvement and chiral symmetry for non-abelian gauge theories at strong coupling.

\section{Improved and unimproved actions}

\begin{figure}[!ht]
 \centering
 \subfigure[\label{fig:imp-comparison}]%
 {\includegraphics[width=0.49\columnwidth]{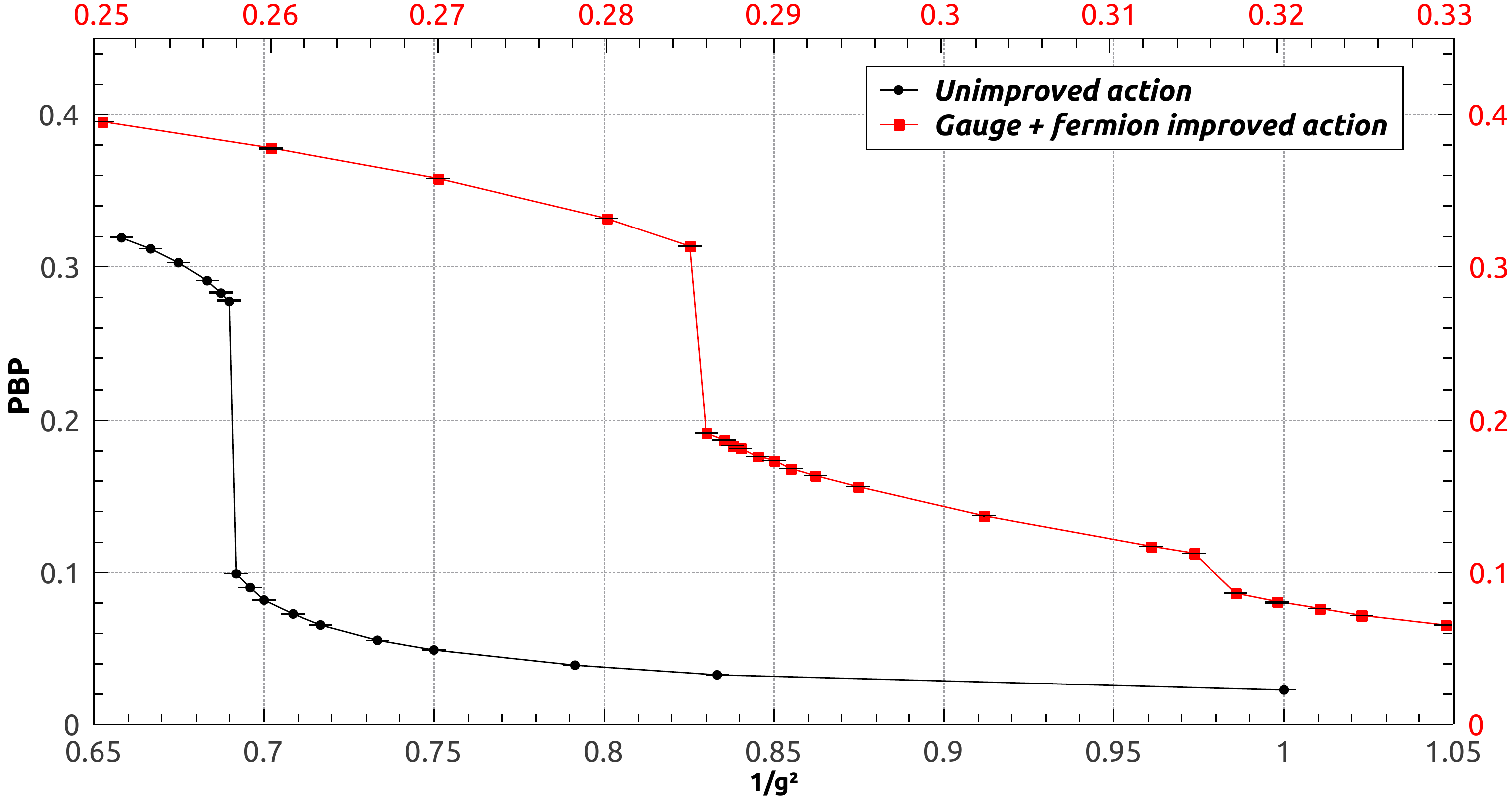}}
 \subfigure[\label{fig:step-imp-comparison}]%
 {\includegraphics[width=0.49\columnwidth]{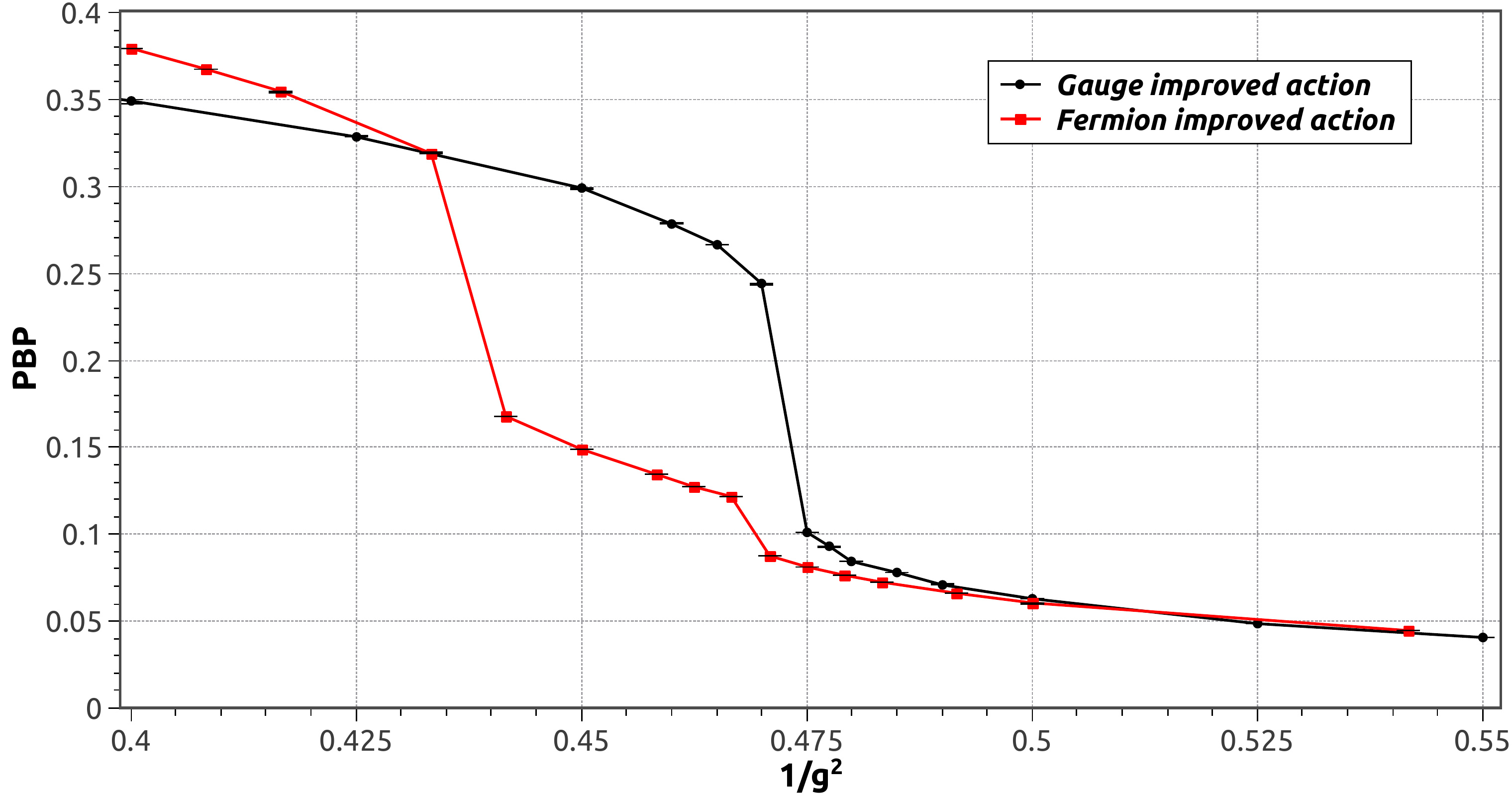}}
 \subfigure[\label{fig:chi-conn-comparison}]%
 %{\includegraphics[width=0.49\columnwidth]{comparison-chiConn-imp-naive.pdf}}
 {\includegraphics[width=0.49\columnwidth]{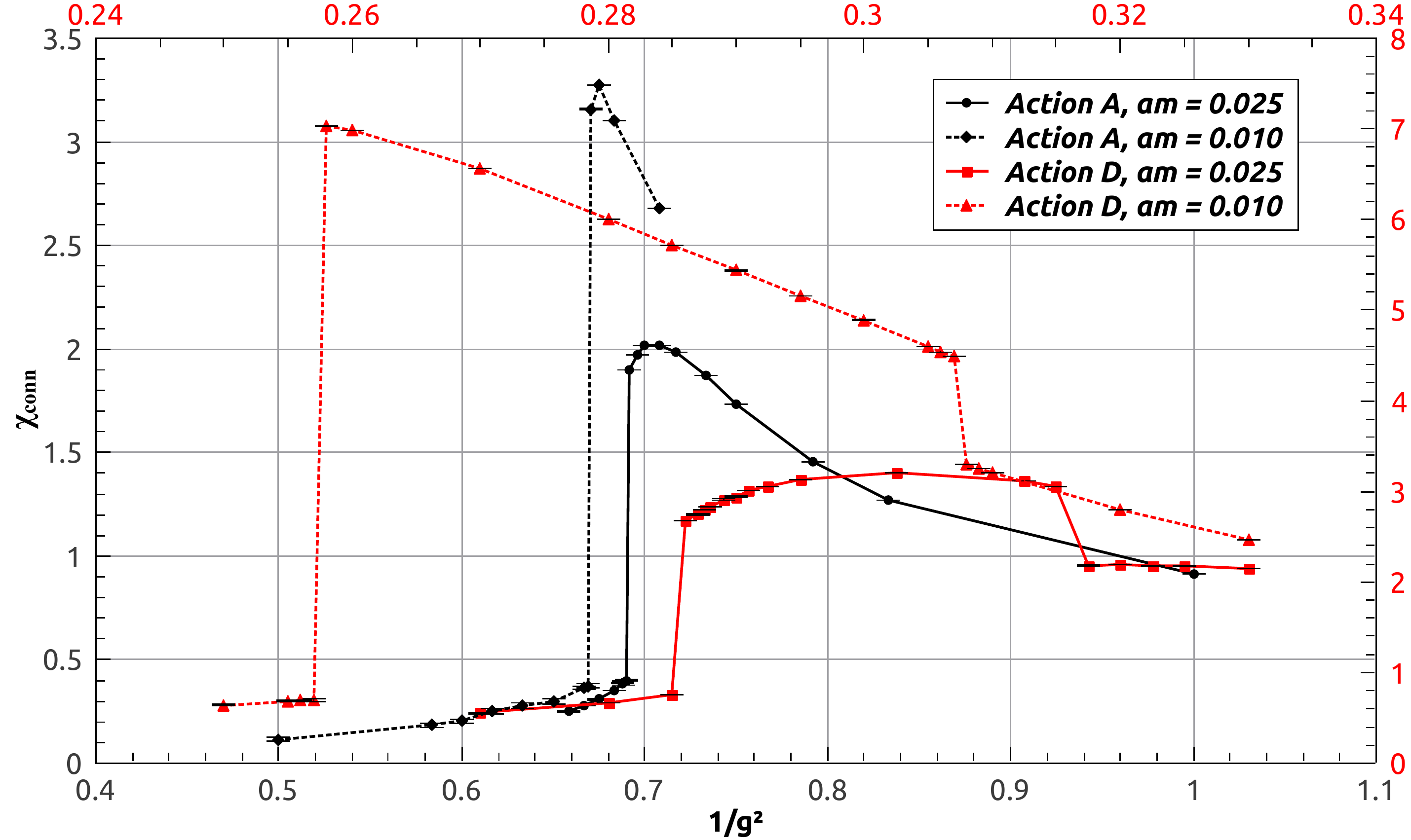}} 
 \subfigure[\label{fig:chi-pi-comparison}]%
 {\includegraphics[width=0.49\columnwidth]{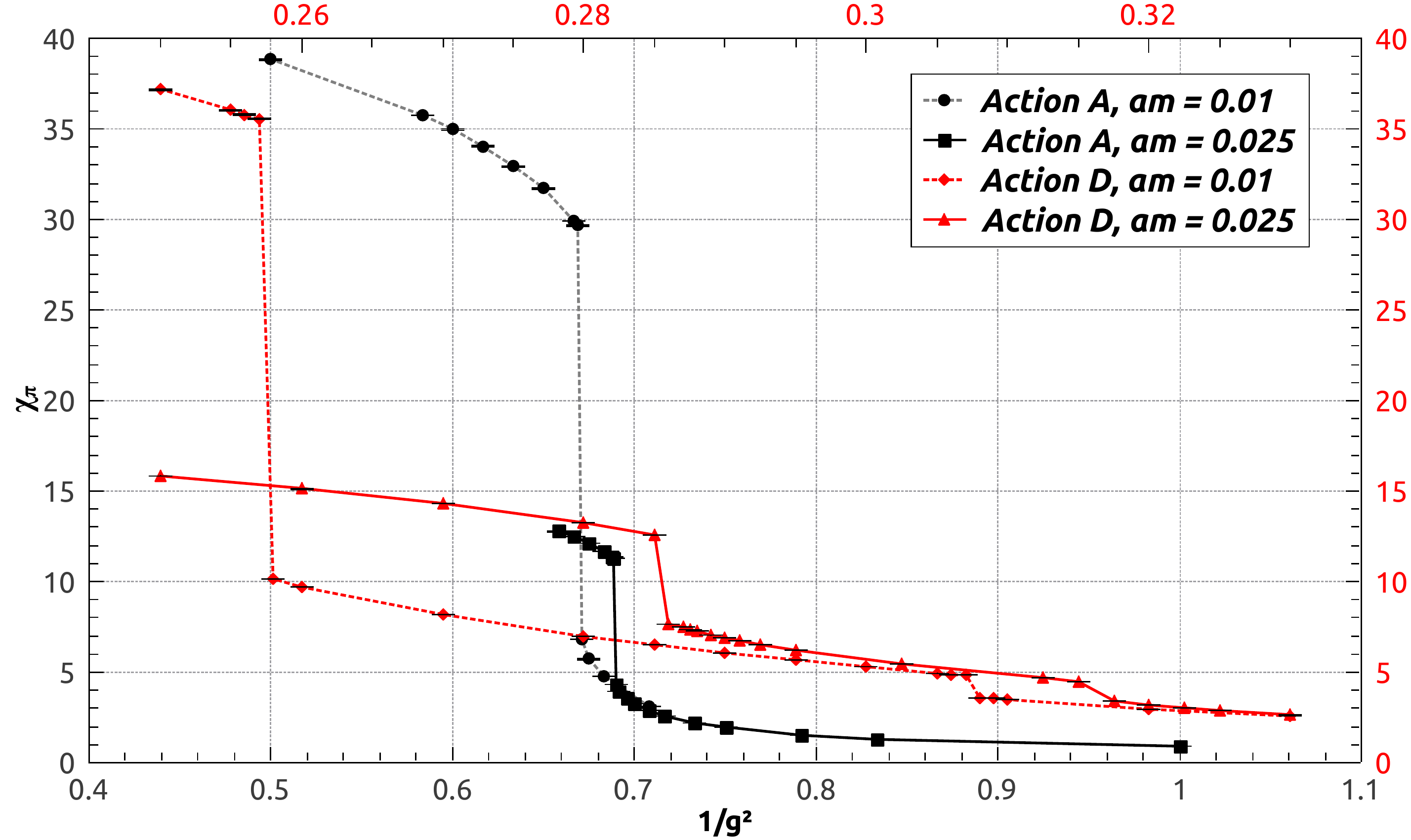}}
 \caption{(a and b) Comparison of the evolution of the chiral condensate for all four actions A to D. The two rapid crossover structure is present for both fermion-improved cases and absent otherwise. (c) Comparison between the connected component of the scalar susceptibility in the unimproved case (Action A) and the improved case (Action D), for two masses. (d) The same for the pseudoscalar susceptibility. (a and d) The unimproved case behaves as expected, while the improved case shows two almost discontinuities.}
 \label{fig:PBP-comparisons}
\end{figure}

In our previous works on $N_f=12$ QCD, we have used a tree-level Symanzik improved gauge action and Naik improved Kogut-Susskind (staggered) fermions. In order to investigate the effect of improvement on the lattice theory at strong coupling, we simulated four different combinations of improved gauge and fermion actions \cite{Deuzeman:2012ee}. These combinations are listed in Table~\ref{tab:actions}. 
\begin{table}[!ht]
\begin{center}
\begin{tabular}{|c | c c |}
\hline
Action & Gauge Improvement & Fermion Improvement \\
\hline
A  &No & No \\
B  & Yes & No \\
C  & No & Yes \\
D  & Yes&  Yes \\
\hline
\end{tabular}
\caption{Four combinations of improved gauge and/or fermion action. Gauge improvement refers to tree-level Symanzik improvement, while fermion improvement refers in this case to the addition of the Naik term to the staggered fermion action. Details can be found in \cite{Deuzeman:2012ee}. }
\label{tab:actions}
\end{center}
\end{table}
Most of the data presented here for the chiral condensate and susceptibilities were obtained at volume $V=16^3\times 24$ (previous scaling studies can be found in \cite{Deuzeman:2011pa}), while all correlators have been studied at volumes $V=24^4$ and $V=32^4$.  
The most notable result, when comparing the plots of the chiral condensate (susceptibilities) in Figure~\ref{fig:PBP-comparisons}, is the presence of a single rapid crossover (almost discontinuity) in the cases where the fermion action is not improved, and  the presence of two rapid crossovers (almost discontinuities) whenever the fermion action is improved. This signals the emergence of a new phase between the two rapid  crossovers. 
Most importantly, we have shown \cite{Deuzeman:2012ee} that the rapid crossover in the chiral condensate observed at weaker coupling {\it diminishes} with decreasing masses. The opposite behaviour is true for the corresponding discontinuity of the connected scalar susceptibility and the pseudoscalar susceptibility $\chi_\pi = \langle\bar{\psi}\psi\rangle /m$.
All observations are consistent with the fact that we are indeed looking at two distinct transitions. Only the one at stronger coupling is the true chiral symmetry breaking transition, while at weaker coupling a second transition due to improvement manifests with a change in the slope of the chiral condensate, i.e. a discontinuity in $\chi_\pi$. 
In \cite{Deuzeman:2012ee} we have explored the features of the new intermediate phase, including correlators, and attempted a theoretical interpretation. We further elaborate on this in   section \ref{sec:improvement}. 
\section{The naive action and the order of the transition}
One point of interest when studying chiral symmetry breaking bulk transitions in the context of conformality is to determine the order of the transition. If the bulk transition is second order, this would signal the emergence in the continuum of a second non trivial ultraviolet attractor (UVFP), in addition to the non trivial infrared attractor (IRFP), which characterizes theories inside the conformal window. If instead the transition is first order, no continuum limit exists. However, a UVFP would emerge at the end-point of a line of first order chiral bulk transitions. A possible scenario for the disappearance of the conformal window could then be realized through  the annihilation of a pair of IR and UV fixed points.

In \cite{Deuzeman:2011pa} we reported evidence, obtained using the improved action D, that the chiral bulk transition of QCD with twelve flavors is first order, thus barring the possibility of a UVFP for the considered action, but still allowing for the latter scenario depicted above. Here, we add evidence that the order of the bulk transition for this theory is not modified by improvement, by studying the transition in the unimproved case.  A more complete analysis will appear elsewhere.
\begin{figure}[!ht]
 \centering
 \subfigure[\label{fig:hysteresis}]%
 {\includegraphics[width=0.49\columnwidth]{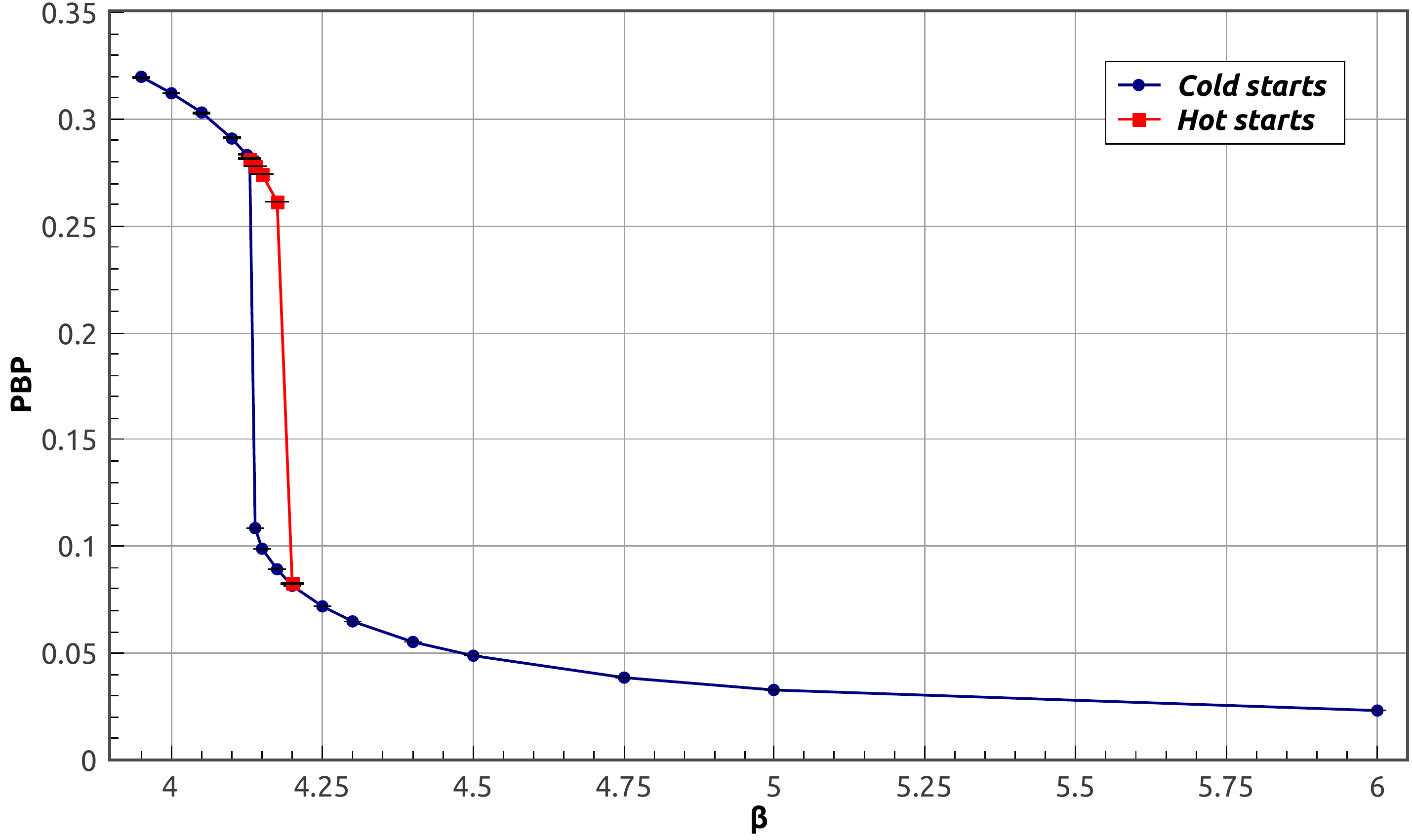}}
 \subfigure[\label{fig:mass-scaling}]%
 {\includegraphics[width=0.49\columnwidth]{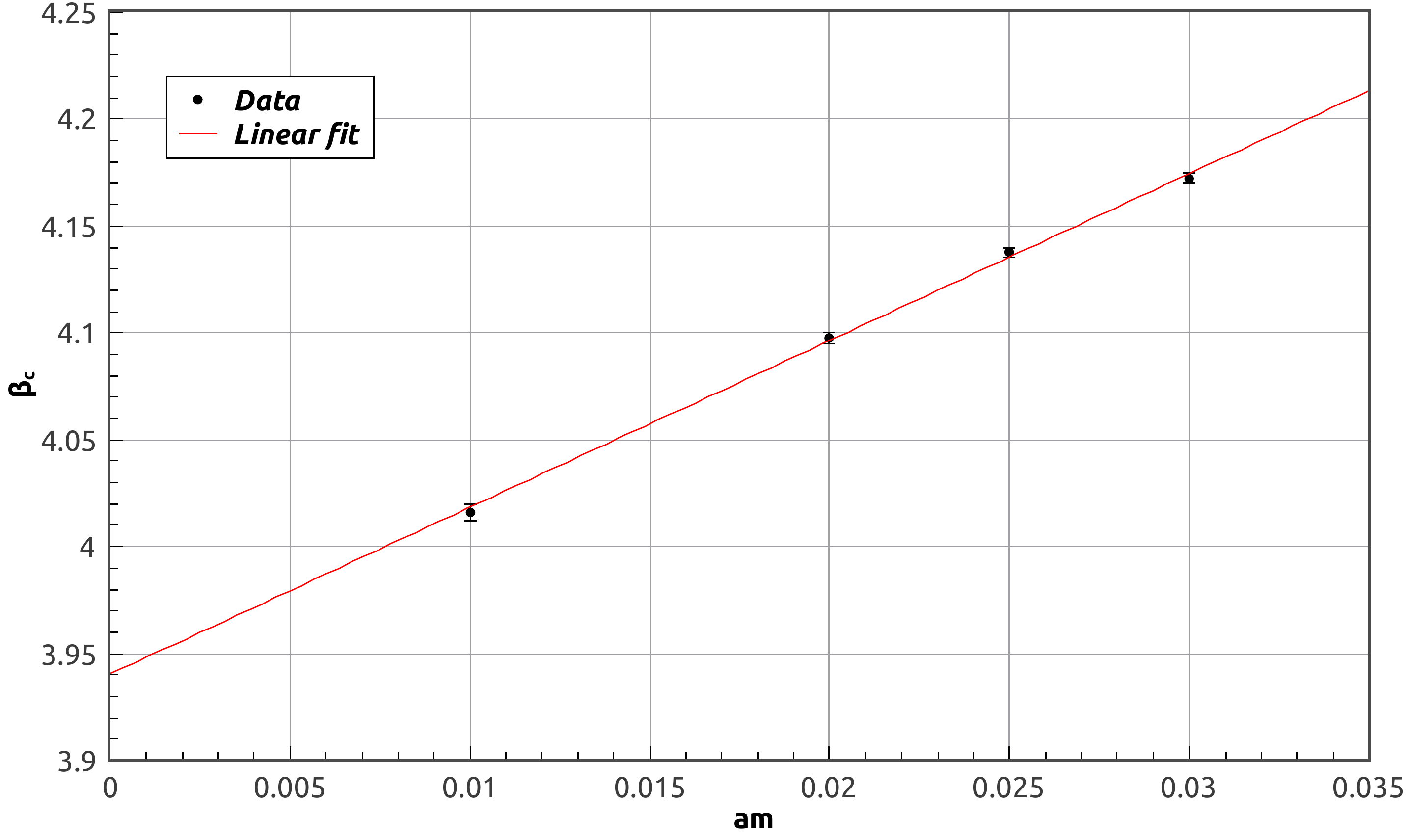}}
 \caption{(a) Hysteresis loop in the chiral condensate obtained from cold (ordered) and hot (disordered) starting lattice configurations; (b) Linear mass scaling of the critical $\beta$ at which the bulk transition takes place for different bare masses; (a and b) Both results are consistent with a first order nature of the transition.}
 \label{fig:first-order}
\end{figure}
In Figure~\ref{fig:hysteresis} we observe a hysteresis loop obtained from runs with different initial conditions. The runs labeled as {\it cold} were started from ordered lattice configurations, while the runs labeled as {\it hot}  were started from disordered lattice configurations, in such a way that they should approach thermalization, respectively, from below and from above. Around a first order transition the existence of metastable states is expected and the runs do not thermalize at the same value for a large number of trajectories, giving rise to the {\it loop} observed in the figure.

Further evidence for a first order nature of this transition comes from the scaling with the mass of $\beta_c = 6/g_c^2$, that is, the critical lattice coupling  at which the bulk transition happens for a given bare mass. We show our results in Figure~\ref{fig:mass-scaling}. They are consistent with the expected linear scaling and fitting with the functional form $y=Ax+b$, yields $A=7.8\pm 0.2$ and $B=3.940\pm 0.005$. %with $\chi^2/d.o.f = 0.899$.

The existence of the hysteresis loop and the linear mass scaling of $\beta_c$ point towards a first order nature for this transition, in agreement with out previous result obtained with the improved action and reported in \cite{Deuzeman:2011pa}. As noticed also in the improved case, we do not observe tunneling between metastable states in the same Monte Carlo history, over a long Monte Carlo time. This could explain the almost discontinuity in the connected scalar  susceptibility shown in Figure \ref{fig:PBP-comparisons}(c).

\section{The effects of improvement, inside and below the conformal window}
\label{sec:improvement}
\begin{figure}[!ht]
 \centering
 \includegraphics[width=0.9999\linewidth]{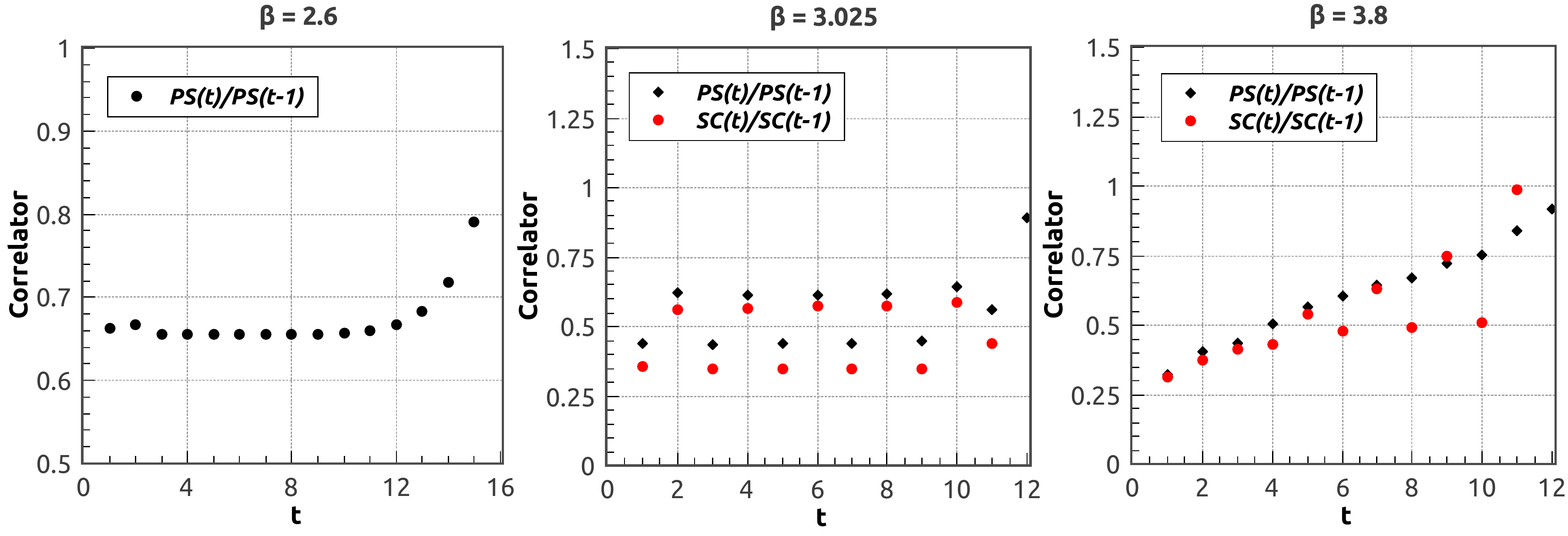}
 \caption{Central values of the ratios $C(t)/C(t-1)$ for the pseudoscalar (PS, black circles) and scalar (SC, red squares) two-point correlation functions for couplings $\beta =10/g^2$ (for action D in the three regions of interest. From left to right: the chirally broken phase, the intermediate phase between the two rapid crossovers and the weak coupling phase. Chiral symmetry is restored in the last two regions.}
 \label{fig:oscillation}
\end{figure}

The spectrum of the theory provides further insight into the effects of improvement of a lattice action at strong coupling. In other words, it allows to identify the properties of the new intermediate phase arising from improvement. We briefly summarize our findings reported in \cite{Deuzeman:2012ee}, and we further comment on the interplay of chiral symmetry and improvement. 
In the $N_f=12$ theory with the improved staggered action D, 
the comparison of the scalar and pseudoscalar staggered two-point functions highlights two effects proper of the intermediate phase: the presence of an oscillatory component in the pseudoscalar correlator and the presence of a forward-backward asymmetry in all correlators. The first is illustrated in Figure~\ref{fig:oscillation}, while the second is discussed in \cite{Deuzeman:2012ee}. 
At weak coupling, the scalar and pseudoscalar correlators are increasingly close to each other, since chiral symmetry is exact and degeneracy should be recovered; as proper of staggered fermions, an oscillating component is present only in the scalar correlator. The growing trend of the ratios is a sign of significant contributions from excited states. At the strongest coupling, chiral symmetry is broken. The pseudoscalar lightest state, being its Goldstone boson, becomes very light and largely non degenerate with the scalar state and has no oscillatory component.

Inside the exotic intermediate phase, the pseudoscalar correlator develops an oscillatory component. This can be explained \cite{Deuzeman:2012ee} by the competition of local and less local contributions to the baryon number density. The less local contributions are generated by improvement; in our case, the most effective term introduced by improvement appears to be the Naik term in the staggered action. 

The way improvement acts at sufficiently strong coupling can be considered analogous to the way the physics of nearest-neighbor spin systems is modified by the presence of non-nearest neighbor interactions. The simplest and instructive example is the Ising spin chain (1D) with a next-to-nearest neighbor term \cite{Arisue:1983vu}. In general, the competition of local and less-local interactions in the theory can generate new phases in the system, as known for ANNNI models \cite{ANNNIReview}.
 
The Symanzik improvement procedure for a gauge or a fermion lattice action is in this respect fully analogous to spin systems. Actually, any improvement - the traditional Symanzik improvement or any  smearing procedure (fattening of links) - explicitly modifies quark and/or gluon interactions carrying higher powers of the lattice spacing. Thus, by construction, improvement accelerates the convergence to the continuum limit for sufficiently small lattice spacings, while it drastically modifies the system at coarse lattice spacings. 
In \cite{Luscher1984} it was explicitly demonstrated that the Transfer matrix of a Symanzik improved lattice gauge action is no longer hermitian, thus complex energy eigenvalues appear and the corresponding amplitudes have an exponentially decaying oscillation. These effects decouple only in the continuum limit, while are increasingly relevant at increasingly coarse lattices. 
New phases can thus appear at strong coupling. Where in the parameter space the new phase is located will depend on the  details of the adopted improvement procedure. 

It is thus not a suprise that this new intermediate phase has been first observed by us in \cite{Deuzeman2010a}  using action D, and subsequently by \cite{Cheng:2011ic} using a n-HYP smeared staggered fermion action and Symanzik improved gauge action. It is also important to emphasize that the broken $S_4$ symmetry mentioned in \cite{Cheng:2011ic} is directly related to the Transfer matrix of staggered fermions via $T=S_4^2$ - this relation can in principle be  applied to all Euclidean directions. A HISQ action, containing an even more aggressive improvement, could enhance these effects when exploring the system at strong coupling.
It thus seems plausible to conclude that {\it the non-hermiticity due to improvement of the Transfer matrix of a lattice action can in general originate new phases of the system for sufficiently strong couplings and coarse lattices}. 

These results can be equally relevant for strongly coupled non-abelian gauge theories inside the conformal window, for example the $N_f=12$ theory studied here, as for strongly coupled systems such as graphene, or non-abelian gauge theories slightly below the conformal window.
When approaching the conformal window from below, the critical temperature at which chiral symmetry is restored decreases with increasing the number of flavors. In other words, the system studied on the lattice will show a thermal transition at increasingly strong lattice coupling for a fixed temporal extent $N_t$: $T_c=1/N_t a(\beta_c)$. 
Thus, we may expect that the same sequence of phases as in Figure 1(a) (red line, improved case) will appear at sufficiently small lattice fermion masses, where now the strongest coupling rapid crossover is the finite temperature chiral symmetry breaking transition, and should show thermal scaling behaviour. 
The chiral symmetry breaking transition would be preceded by a new phase due to improvement. The additional crossover at weaker coupling would not signal chiral symmetry breaking  (the crossover disappears in the chiral limit), but only a change of slope in the mass dependence of the chiral condensate. As also shown in \cite{Deuzeman:2011pa}, this transition can acquire a peculiar $N_t$ dependence, disjoint from the physics of the true strong coupling chiral  transition. 
In this situation, at such coarse lattices and small masses, the analysis of the physical properties of the system inside or below the conformal window clearly becomes more complex. For fundamental fermions the guiding observable should remain the chiral condensate, the only true order parameter in the system. Many other observables, including Dirac operator eigenvalues, will be unavoidably affected by large lattice artifacts, and decreasing fermion masses will push the system into the small volume (usually referred to as $\epsilon$ and $\epsilon'$) regime. 
\section{Conclusions and outlook}
After summarizing the results recently reported in Ref.~\cite{Deuzeman:2012ee},
 we have added evidence that the chiral symmetry breaking bulk transition for the $N_f=12$ theory with the non-improved action is of first-order nature, in full agreement with our previous findings. A UVFP can thus appear at the end-point of a line of first order bulk transitions. It is presently unknown, but worth investigating, if this end-point plays a role in the disappearance of the conformal window. 

As shown in \cite{Deuzeman:2012ee}, a new intermediate phase at weaker coupling precedes the chiral symmetry breaking transition, in the presence of improvement.  We stress that 
this is a general feature of improvement, while the region of the parameter space where the new phase occurs depends on the details of the improvement procedure.
The emergence of this new phase, its causes and properties, can be equally relevant for the study of non-abelian gauge theories in the conformal window, or slightly below it, and the lattice simulation of strongly coupled systems such as graphene.

\section*{Acknowledgments}
We warmly thank our collaborators A. Deuzeman, M.P. Lombardo and K. Miura for the always pleasant and inspiring work together. 
This work was in part based on the MILC collaboration's public lattice gauge theory code. Computer time was provided through the Dutch National Computing Foundation (NCF) and the University of Groningen.

\end{document}